\documentclass[twocolumn,showpacs,preprintnumbers,amsmath,amssymb]{revtex4}

\usepackage{graphicx}
\usepackage{dcolumn}
\usepackage{bm}

\begin{document}

\title{Bond-bending modes and stability of tetrahedral semiconductors under high pressure:
a puzzle of AlN}

\author{Ekaterina V. Iakovenko$^a$}
\author{Michel Gauthier$^b$}
\author{Alain Polian$^b$}%
\affiliation{%
$^a$Institute for High Pressure Physics, Russian Academy of
Sciences 142190 Troitsk, Moscow Region, Russia\\
$^b$Physique des Milieux Condens\'{e}s, Universit\'{e} P\&M Curie, F 75252 Paris Cedex 05, France
}%

\date{\today}

\begin{abstract}
Lattice vibrations of the w\"{u}rtzite-type AlN have been studied by Raman spectroscopy
under high pressure up to the structural phase transition at $20$ GPa. We have shown that
the widely debated bond-bending $E_2^1$ mode of w-AlN has an abnormal positive pressure
shift up to the threshold of the phase transition,
whereas in many tetrahedral semiconductors the bond-bending modes soften on compression.
 This finding disagrees with the results of
{\it ab initio} calculations, which give a "normal" negative pressure shift.
 Combination of high {\it dynamical} and  low {\it thermodynamical} stability of AlN
breaks the correlation between the mode Gr\"{u}neisen parameters for the bond-bending modes and
the transition pressure, which holds for CdS, InP, ZnO, ZnTe, ZnSe, ZnS, Ge, Si, GaP, GaN, SiC
 and BeO.
\end{abstract}

\maketitle

For the last four decades, a rather curious and at first sight surprising phenomenon
of softening of tetrahedral semiconductors under compression has been reported in a number of
experimental \cite{payne, weinstein1, weinstein2, weinstein3, olego, weinstein4,
yakovenko, perlin, olijnik, klotz} and theoretical \cite{dolling, martin, jex, wendel, yin,
chang, nielsen, chadi} works. Thus, Si, Ge, A$_{\rm III}$B$_{\rm V}$ and A$_{\rm II}$B$_{\rm VI}$
semiconductors have
pressure-sensitive phonon modes with negative pressure shifts \cite{weinstein2,
yakovenko,perlin,klotz}. These soft modes are shearing modes, involving bond bending
in the first order of the strain \cite{yin,chadi}. Phonon frequency drop, more pronounced
for high-Z materials, reaches about $30$\% at the threshold of the pressure induced
phase transitions, when covalent
tetrahedral structures lose their stability and transform to more densely packed arrangements.

Earlier, the only known example of a bond-bending mode having a positive, although weak,
pressure shift was being the "pure bread" bond-bending $E_2^1$ mode of the low-Z w\"{u}rtzite-type
BeO \cite{jephcoat}. This result was being regarded as an exception until recent Raman
measurements found a similar behavior for the $E_2^1$ mode of the w\"{u}rtzite-type AlN
\cite{goni}. {\it Ab initio} calculations \cite{ goni,gorczyca} did not catch this feature,
 giving a "normal" negative pressure shift for the $E_2^1$ mode of AlN. This discrepancy
 deserves special attention, since the bond-bending elasticity
is one of the most prominent manifestations of directional covalent bonding, and its
pressure behavior should be well accounted for. The point to be made here is that AlN
represents a complicated case of covalent {\it versus} ionic bonding \cite{karch}:
 although its valence charge distribution is highly io nic \cite{gabe},
AlN adopts the tetrahedrally coordinated w\"{u}rtzite structure under ambient
conditions \cite{harrison}.
 Aspiring to clarify the question, we have undertaken
 a complementary high-pressure Raman study of vibrational modes of the w\"{u}rtzite-type AlN
up to its stability limit at about $20$ GPa. The pressure dependence of the low-frequency
bond-bending $E_2^1$ mode is traced up to the threshold of the pressure induced phase transition
 for the first time.

The AlN samples were $20$ $\mu$m-thick crystals grown on the sapphire substrate by vapor
phase epitaxy. Pressure was
produced using the diamond-anvil pressure cell. Compressed helium and methanol-ethanol
mixture were used as a pressure-transmitting medium in the first and in the second
experimental run, respectively. Pressure was measured {\it in situ} by the ruby luminescence
technique. The Raman spectra were measured using the THR-$1000$ triple spectrometer equipped
with an OSMA detector (the first run), and the Dilor XY double spectrometer equipped with
 the CCD detector (the second run). An Ar$^+$ laser ($\lambda=514.5$ nm) was used as a source of
excitation. All spectra were recorded in the backscattering geometry at ambient temperature.

For the hexagonal w\"{u}rtzite structure with the space group $P6_3$mc ($Z=2$), a factor-group
 analysis predicts the following six sets of optical modes at the ${\bf k}=0$ \cite{fateley}:
 \[\Gamma_{op} = A_1+2B_1+E_1+2E_2, \]
 where $A_1$, $E_1$ and $E_2$ are Raman active modes, and $B_1$ modes are silent. $A_1$
 and $E_1$ are also
infrared active, and split into the longitudinal and transverse components (LO and TO).
The lowest-frequency $E_2^1$ mode is a bond-bending mode.

\begin{figure}
\includegraphics{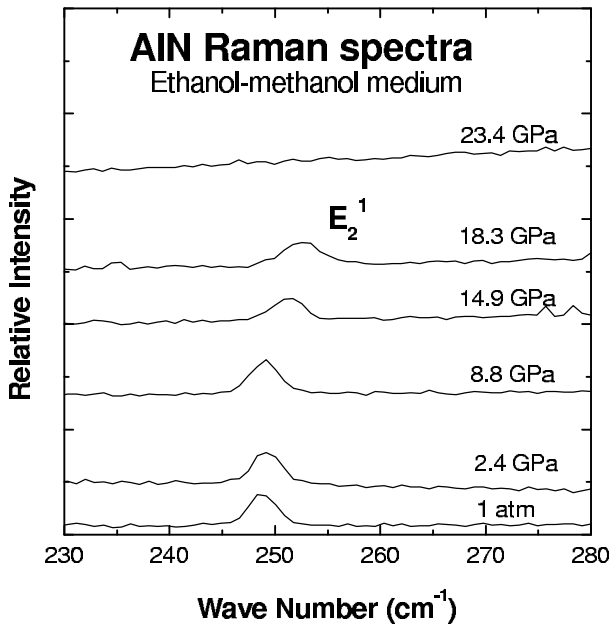}%
\caption{\label{fig:fig1} Raman spectra of AlN as a function of pressure in the
low-energy region. The spectral resolution is $0.5$ cm$^{-1}$.}
\end{figure}

The Raman spectrum of w-AlN has been measured previously under ambient conditions and analyzed
 in some detail, including the effects of polarization and anisotropy
 \cite{mcneil,filippidis,davydov}.
Our ambient pressure Raman frequencies
$\nu_{E_2^1}=249$ cm$^{-1}$,
$\nu_{A_1(TO)}=610$ cm$^{-1}$,
$\nu_{E_2^2}=657$ cm$^{-1}$,
$\nu_{E_1(TO)}=669$ cm$^{-1}$,
$\nu_{A_1(LO)}=890$ cm$^{-1}$,
and $\nu_{E_1(LO)}=910$ cm$^{-1}$
agree with very reliable data of Ref.\cite{mcneil,filippidis,davydov} within $1\%$.
On increase in pressure, all Raman bands shift continuously
to higher phonon energy  with no intensity loss
to about $18$ GPa. Above $18$ GPa, the bands weaken and disappear at about $21$ GPa in both experimental runs due to the
phase transition to the rock salt structure \cite{ueno,xia}.
Representative Raman spectra of w-AlN in the low-energy region as a function
of pressure are shown in Fig.~\ref{fig:fig1}.

\begin{figure}
\includegraphics{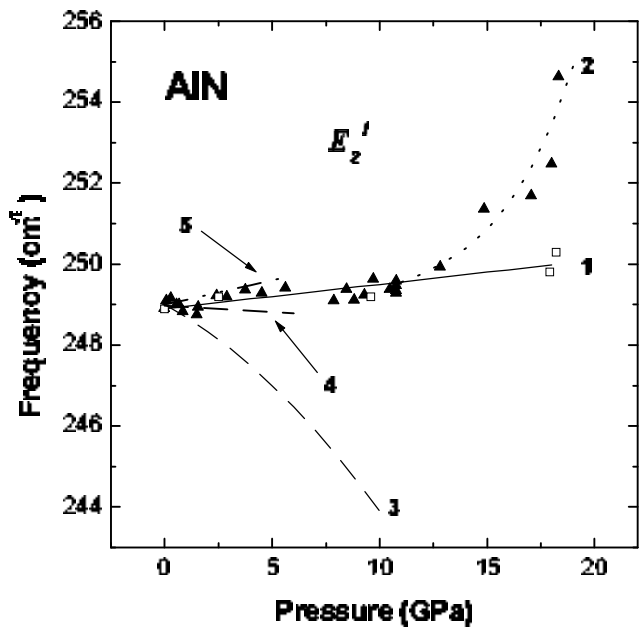}%
\caption{\label{fig:fig3} Comparison between the measured and calculated pressure
dependence of the Raman frequency for the $E_2^1$ mode. The squares are
the first run data, obtained with compressed helium as a pressure transmitting
medium. The triangles are the second run data, obtained with ethanol-methanol
mixture as a pressure transmitting medium. The solid line \textbf{1} is a linear fit of the
first run data. The dotted line \textbf{2} is guided for the eye. The lines \textbf{3}
 and \textbf{4} are calculated
dependences obtained in \cite{gorczyca} and \cite{ goni}, respectively. The line \textbf{5}
is the experimental dependence
obtained in \cite{ goni}. All data are shifted along the vertical axis in order to coincide
 the $1$ atm frequencies with the value $249$ cm$^{-1}$ obtained in our first experimental run.}
\end{figure}

Fig.~\ref{fig:fig3} compares measured and calculated pressure
dependences of the Raman frequency for the $E_2^1$ mode. The mode pressure
coefficients $\nu_i'$ calculated using the linear least-squares fit
$\nu_i = \nu_i^0 + \nu_i'P$, where $\nu_i$ is the frequency of the mode $i$ at pressure $P$,
 are listed in Table I \cite{comment1}.
The pressure dependence of the $E_2^1$ frequency, measured in our experiment,
is weak but apparently positive and linear at pressures to the
threshold of the phase transition \cite{comment4}.
This finding confirms the results obtained in \cite{goni}, although
our pressure slope of the $E_2^1$ frequency is somewhat lower (see
Fig.~\ref{fig:fig3}). So, the positive pressure shift of the $E_2^1$
frequency of AlN should be regarded as a reliable experimental fact.

As we have mentioned above, {\it ab initio} calculations \cite{ goni,gorczyca}
do not reproduce this feature of AlN, although the agreement
between measured and calculated  pressure shifts for the
high-frequency bond-stretching
$E_2^2$, TO and LO modes is very good (see Table I).
Certainly, the value of the discussed effect for the $E_2^1$ mode
is quite small, and the lack of {\it ab initio} calculations is
not so crucial. Nevertheless, this lack is very instructive to
realize a subtle pressure-sensitive balance between stabilizing and destabilizing
contributions to the bond-bending force constant \cite{wendel,yin,martin}
and may serve a further improvement of the theory.

\begin{table}
\caption{\label{tab:table1}Measured and calculated linear pressure coefficients $\nu_i'$
(cm$^{-1}$GPa$^{-1}$) for the AlN Raman frequencies.}
\begin{ruledtabular}
\begin{tabular}{lcccccc}
 &$E_2^1$ &$A_1{(TO)}$ &$E_2^2$&
 $E_1{(TO)}$ & $A_1{(LO)}$ &$E_1{(LO)}$\\
\hline
\textbf{Expt.} & & & & & & \\
Run 1 & 0.05(1) & 3.8(2) & 4.9(2) & 4.5(1) & - & -\\
Run 2 & 0.05(1) & 4.3(2) & 4.65(3) & 4.55(6) & 4.0(1) & 3.6(7)\\
Go\~{n}i\footnotemark[1] & 0.12(5) & 4.4(1) & 4.99(3) & 4.55(3) & - & 4.61(1)\\
\textbf{Calc.} & & & & & &\\
Gorczyca\footnotemark[2] & -0.29 & 4.29 & 4.79 & 4.36 & - & -\\
Go\~{n}i\footnotemark[1] & -0.03 & 3.0 & 4.2 & 3.8 & 3.5 & 4.0
\end{tabular}
\end{ruledtabular}
\footnotetext[1]{Ref.~\onlinecite{goni}.}
\footnotetext[2]{Ref.~\onlinecite{gorczyca}.}
\end{table}

Table II compiles the commonly used mode-Gr\"{u}neisen parameters
$\gamma_i = -d\ln\nu_i/d\ln V$ obtained
 for the bond-bending modes $i$ in a series of tetrahedral compounds \cite{comment2}. Negative
value for $\gamma_i$
is observed in each case, except for SiC, BeO and AlN. Although SiC has an essentially
zero $\gamma_i$, its quadratic pressure coefficient is negative \cite{yakovenko}.
Thus, BeO and AlN appear to be the most stable materials with respect to the bond-bending mode
on compression.

\begin{table}
\caption{\label{tab:table2}Mode-Gr\"{u}neisen parameters for
the bond-bending modes.}
\begin{ruledtabular}
\begin{tabular}{llcl}
Material & Structure &Bond-bending mode & $\gamma_i$\\
\hline
CdS & w\"{u}rtzite & $E_2$ & -2.7\footnotemark[1] \\
InP & zinc blende & TA(L) & -2.0\footnotemark[1] \\
ZnO & w\"{u}rtzite&$E_2$ & -1.8\footnotemark[1]  \\
GaAs &zinc blende &TA(L) & -1.7\footnotemark[1]  \\
ZnTe &zinc blende &TA(L) & -1.5\footnotemark[1]  \\
Ge & diamond &TA(L) & -1.52\footnotemark[2]  \\
Si &diamond & TA(L)& -1.3\footnotemark[3]   \\
ZnS & zinc blende & TA(L) & -1.18\footnotemark[1]   \\
GaP & zinc blende & TA(L) & -0.81\footnotemark[4]  \\
GaN & w\"{u}rtzite& $E_2$ & -0.426\footnotemark[5]  \\
SiC & hex.(6H) & $E_2$& 0.0\footnotemark[6]   \\
BeO &w\"{u}rtzite &$E_2$ & 0.04\footnotemark[7]   \\
AlN &w\"{u}rtzite &$E_2$ & 0.04\footnotemark[8]  \\
\end{tabular}
\end{ruledtabular}
\footnotetext[1]{Ref.~\onlinecite{weinstein4}.}
\footnotetext[2]{Ref.~\onlinecite{klotz}.}
\footnotetext[3]{Ref.~\onlinecite{weinstein2}.}
\footnotetext[4]{Ref.~\onlinecite{weinstein1}.}
\footnotetext[5]{Ref.~\onlinecite{perlin}.}
\footnotetext[6]{Ref.~\onlinecite{yakovenko}.}
\footnotetext[7]{Ref.~\onlinecite{jephcoat}.}
\footnotetext[8]{This study.}
\end{table}

This bond-bending stability of AlN looks rather surprising, since its w\"{u}rtzite phase has
 a rather limited stability range up to $21$ GPa. Indeed, experimentally it
has been found that the stability of tetrahedral structures with respect
to the bond-bending modes correlates with their absolute stability under pressure
in such a way that the frequency drop is faster for less stable compounds.
Weinstein \cite{weinstein3,weinstein4} has discovered that for six
diamond and zinc-blende structure semiconductors ZnTe, Ge, Si, ZnSe, ZnS and GaP there
is a remarkable linearity between the mode Gr\"{u}neisen parameter $\gamma_{TA(X)}$ for the
bond-bending TA(X) mode and the transition pressure $P_{tr}$ for these materials.
{\it Ab initio} calculations by Yin and Cohen \cite{yin} have given some insight into
the nature of this correlation. They have shown that those individual contributions to
the total crystal energy of Si and Ge, which stabilize the bond-bending TA(X) mode,
are just the same ones, which preserve the diamond structure from rearrangement to more
densely packed structures \cite{comment3}. The balance between stabilizing and destabilizing
contributions drastically depends on the specific volume, destabilizing contributions
 becoming stronger under compression.

\begin{figure}
\includegraphics{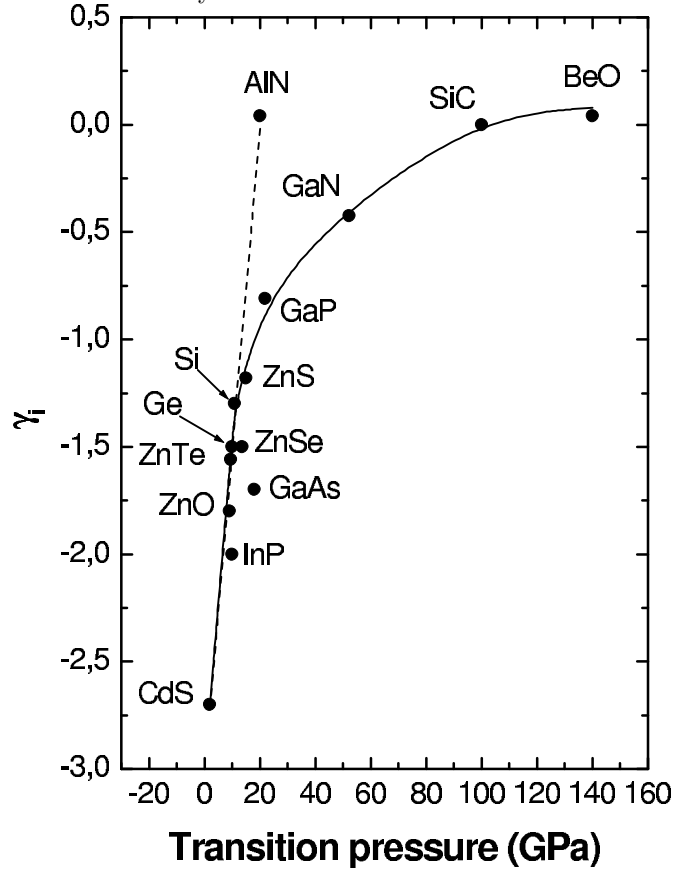}%
\caption{\label{fig:fig5} Correlation between the mode Gr\"{u}neisen parameter $\gamma_i$
for the bond-bending TA(L) [$E_2$] modes and the transition pressure in a series
of tetrahedral semiconductors. Solid and dashed lines are guided for the eye.
References are given in Table II and in the text.}
\end{figure}

At present it is possible, following Weinstein and Zallen \cite{weinstein3,weinstein4}, to trace
the $\gamma_i - P_{tr}$ correlation to much higher pressures, using recent data
for w-GaN \cite{perlin}, SiC-$6H$ \cite{yakovenko,yoshida} and w-BeO \cite{mori}. The mode Gr\"{u}neisen
 parameters determined for the bond-bending TA(L) phonons of InP, GaAs, ZnTe, ZnS, Ge, Si,
GaP and for the bond-bending $E_2$ phonons of CdS, ZnO, AlN, GaN, SiC and BeO are plotted in
Fig.~\ref{fig:fig5} as a function of the transition pressure $P_{tr}$ ranging to $140$
GPa.

 Fig.~\ref{fig:fig5} shows that the $\gamma_i$ {\it versus}  $P_{tr}$ points form a smooth curve
deviating from linearity at ${\rm P}_{tr}>20$ GPa
and approaching to zero at $P_{tr}\approx 100$ GPa. On the basis of this curve
one may reasonably
expect that materials with $\gamma_i > 0$ should be at least as stable under
pressure as SiC does.
However, the point for AlN falls far afield from the common curve and tends
to continue a linear dependence drawn by high-Z materials.

So, we see that
a simple universal correlation between the pretransitional behavior of
the bond-bending modes and the stability
 of tetrahedral semiconductors under high pressure does not exist. To our knowledge,
 the question why AlN combines a high {\it dynamical} and a low {\it thermodynamical} stability remains
 still open, and a detailed microscopic treatment,
 possibly along the direction indicated by {\it Yin} and {\it Cohen} \cite{yin},
 is required.

The authors wish to thank A. Dobrynin for growing the AlN crystals. E. V. Iakovenko is
grateful to A. F. Goncharov for his assistance in Raman measurements.


\begin{thebibliography}{99}

\bibitem{payne} R. T. Payne, Phys. Rev. Lett. \textbf{13}, 53 (1964).

\bibitem{weinstein1} B. A. Weinstein and G. J. Piermarini, Phys. Lett. A
\textbf{48}, 14 (1974).

\bibitem{weinstein2} B. A. Weinstein and G. J. Piermarini, Phys. Rev. B \textbf{12},
 1172 (1975).

\bibitem{weinstein3} B. A. Weinstein, Solid State Commun. \textbf{24}, 595 (1977).

\bibitem{olego} D. Olego and M. Cardona, Phys.Rev. B \textbf{25}, 1151 (1982).

\bibitem{weinstein4} B. A. Weinstein and R. Zallen, in {\it Light Scattering in Solids IV},
 edited by M. Cardona and G. Guntherodt (Springer, Heidelberg, 1984).

\bibitem{yakovenko} E. V. Yakovenko, A. F. Goncharov, S. M. Stishov, High Pressure Research
\textbf{7}, 433 (1991).

\bibitem{perlin} P. Perlin, C. Jauberthie-Carillon, J. P. Itie, A. San Miguel,
I. Grzegory, and A. Polian, Phys. Rev. B \textbf{45}, 83 (1992).

\bibitem{olijnik} H. Olijnyk, High Pressure Research \textbf{10}, 461 (1992).

\bibitem{klotz} S. Klotz, J. M. Besson, M. Braden, K. Karch, P. Pavone, D. Strauch,
and W. G. Marshall, Phys. Rev. Lett. \textbf{79}, 1313 (1997).

\bibitem{dolling} G. Dolling, R. A. Cowley, Proc. Phys. Soc. \textbf{88}, 463 (1966).

\bibitem{martin} R. M. Martin, Phys. Rev. \textbf{186}, 871 (1969).

\bibitem{jex} H. Jex, Phys. Status Solidi (b) \textbf{45}, 343 (1971).

\bibitem{wendel} H. Wendel and R. M. Martin, Phys. Rev. B \textbf{19}, 5251 (1979).

\bibitem{yin} M. T. Yin and M. L. Cohen, Phys. Rev. B \textbf{26}, 3259 (1982); \textbf{26},
 5668 (1982).

\bibitem{chang} K. J. Chang and M. L. Cohen, Phys. Rev. B \textbf{31}, 7819 (1985);
 \textbf{34}, 8581 (1986).

\bibitem{nielsen} O. H. Nielsen, Phys. Rev. B \textbf{34}, 5808 (1986).

\bibitem{chadi} D. J. Chadi and R.M. Martin, Solid State Communications \textbf{19},
643 (1976).

\bibitem{jephcoat} A. P. Jephcoat, R. J. Hemley, H. K. Mao, R.E. Cohen, and M.J. Mehl,
Phys. Rev. B \textbf{37}, 4727 (1988).

\bibitem{goni} A. R. Go\~{n}i, H. Siegle, K. Syassen, C. Thomsen, and J.-M. Wagner,
Phys. Rev. B \textbf{64}, 035205 (2001).

\bibitem{gorczyca} I. Gorczyca, N. E. Cristensen, E. L. Peltzer y Blanca,
and C. O. Rodriguez, Phys. Rev. B \textbf{51}, 11936 (1995).

\bibitem{karch} K. Karch and F. Bechstedt, Phys. Rev. B \textbf{56}, 7404 (1997).

\bibitem{gabe} E. Gabe, Y. LePage, and S. L. Mair, Phys. Rev. B \textbf{24}, 5634 (1981).

\bibitem{harrison} W. Harrison, {\it Electronic structure and Properties of Solids}
(Freeman, San Francisco, 1980).

\bibitem{fateley} W. G. Fateley, F. R. Dollish, N. T. McDevitt, and F.F Bentley,
{\it Infrared and Raman Selection Rules for Molecular and Lattice Vibrations:
The Correlation Method} (Wiley-Interscience, New York, 1972).

\bibitem{mcneil} L. E. McNeil, M. Grimsditch, and R. H. French,
J. Am. Ceram. Soc. \textbf{76}, 1132 (1993).

\bibitem{filippidis} L. Filippidis, H. Siegle, A. Hoffman, C. Thomsen, K. Karch,
and F. Bechstedt, Phys.Status Solidi (b) \textbf{198}, 621 (1996).

\bibitem{davydov} Yu. Davydov, Yu. E. Kitaev, I. N. Goncharuk, A. N. Smirnov, J. Graul,
O. Semchinova, D. Uffmann, M. B. Smirnov, A. P. Mirgorodsky, and R. A. Evarestov,
Phys. Rev. B \textbf{58}, 12 899 (1998).

\bibitem{ueno} M. Ueno, A. Onodera, O. Shimomura,  K. Takemura, Phys. Rev. B \textbf{45},
 10123 (1992).

\bibitem{xia} Q. Xia, H. Xia, A. L. Ruoff, J. Appl. Phys. \textbf{73}, 8198 (1993).

\bibitem{comment1} The $E_2^1$ frequency in the second run was fitted only to $12$ GPa.

\bibitem{comment4} A sudden rise, observed in the second run at $P>13$ GPa
is obviously assosiated to the solidification of the ethanol-methanol medium,
 resulting in a nonuniform sample stress. So, above $13$ GPa the data obtained in hydrostatic
 conditions with compressed helium as a pressure transmitting medium (the first run)
are the most reliable ones.

\bibitem{comment2} The mode-Gruneisen parameters $\gamma_i = -d\ln\nu_i/d\ln V$
are related at
zero pressure with the coefficients $\nu_i'$ by the equation $\gamma_i = (B_0/\nu_i^0)\nu_i'$,
where
 $\nu_i^0$ is the mode frequency and $B_0$ is the bulk modulus at ambient pressure.

\bibitem{comment3} Namely, the electronic contributions (i.e., the resulting contribution
from the electronic kinetic energy,
 the electron-core interaction energy, the electron-electron Coulomb energy, and
the electronic exchange and correlation energy) favor phases with low coordination numbers
 and stabilize the bond-bending TA(X) mode of the diamond structure, while the core-core
Coulomb energy (the Ewald energy) favors phases with high coordination numbers and
destabilizes the TA(X) mode.

\bibitem{yoshida} M. Yoshida, A. Onodera, M. Ueno, K. Takemura, and O. Shimomura,
Phys. Rev. B \textbf{48}, 10587 (1993).

\bibitem{mori} Y. Mori, T. Ikai, K. Takarabe, Abstracts of High Pressure
Conference of Japan, 27-29 Nov., 2002, printed  in Special Issue of The Review
of High Pressure Science and Technology, volume 12, p. 2D05, 2002 .



\end{thebibliography}

\end{document}